# Coherence-mediated squeezing of cavity field coupled to a coherently driven single quantum dot

Parvendra Kumar[1,2,*] and Agnikumar G. Vedeshwar[1,†]

[1]*Thin Film Laboratory, Department of Physics and Astrophysics, University of Delhi, Delhi 110007, India*
[2]*Department of Physics, Bhagini Nivedita College, University of Delhi, New Delhi-110043, India*
[*]parvendra1986@gmail.com, [†]agni@physics.du.ac.in

Coherence has been remaining a key resource for numerous applications of quantum physics ranging from quantum metrology to quantum information. Here, we report a theoretical work on how maximally created coherence results in the squeezing of cavity field coupled to a coherently driven single quantum dot. We employ a polaron master equation theory for accurately incorporating the impact of exciton-phonon coupling on squeezing.

## I. Introduction

Quantum coherence is an important ubiquitous feature of quantum physics that results from the superposition of constituent states of the quantum system. It has been exploited extensively and serves as an important resource in many diverse fields such as quantum information and quantum computing [1-3], quantum metrology [4-6], quantum biology [7, 8], and quantum transport [9, 10]. Further, the ability to create and control the coherence is a key requirement for the realization of several practically important concepts such as coherent population trapping and electromagnetically induced transparency [11, 16], steady state population inversion [17], and lasing without inversion [18]. Recently, coherence is also shown to facilitate the squeezing of the resonance fluorescence of a two-level atom [19]. It is to be noted that this approach is quite different from the usual approach based on non-linear processes in variety of systems such as Kerr effect and wave mixing in atomic and solid-state systems [20–23], cavity quantum electrodynamics and cavity optomechanics [24-26], and Bose-Einstein condensation [27]. Here, it is quite worthwhile to mention the squeezing that refers to a noise reduction below the shot noise in either of the quadrature but at the expense of increased noise in a canonically conjugate quadrature. Therefore, squeezing is found to be quite useful for various applications in quantum metrology and gravitational wave detection [28-31].

Recently, there has been a great interest to utilize the solid state systems like quantum dots, superconducting circuits, and NV center in diamond for the generation of quantum light, viz. single-photons and entangled photon-pairs [32-35], and squeezed light [36]. In particular, the squeezing of the resonance fluorescence of a single quantum dot has been demonstrated very recently [37-39]. However, for practical applications, squeezing of the cavity mode field seems to be quite useful and desirable too. This is due to the fact that the light emitted via cavity mode can be extracted much efficiently through an optical fiber or a waveguide. In this paper, we theoretically demonstrate the squeezing of cavity field coupled to a coherently driven single quantum dot. Firstly, we investigate the creation of maximum coherence between ground and exciton state of QD and then show that how the squeezing depends upon the created coherence. Subsequently, we went on to investigate the occupation probabilities of various Fock states of the cavity mode and coherences between them for illustrating that the squeezing of cavity field originates from the build-up of exciton coherence which results in the creation of appreciable coherence only between zero- and one-photon Fock states of the cavity mode. Since the exciton state of a quantum dot unavoidably interacts significantly with its phonon modes, we also analyze the impact of exciton-phonon coupling on the squeezing of cavity field by employing a polaron master equation theory based on polaron transformation approach.

## II. Theoretical Formalism
### A. Model of the dot-cavity system

We consider a single self-assembled InGaAs/GaAs quantum dot (QD) coupled to a single-mode pillar microcavity. The QD is modeled as a two-level system consisting its ground and exciton states represented by $|g\rangle$ and $|e\rangle$, respectively. A schematic sketch of the QD excitation by continuous wave (CW) laser field with associated Rabi frequency $\Omega$ and coupling of a cavity mode of associated coupling strength $g_c$ is shown in Fig. 1. The joint state vector $|0,g\rangle$, represents zero-photon Fock state of cavity mode and ground state of QD. Similarly, $|0,e\rangle$, $|1,g\rangle$, $|1,e\rangle$, and $|2,g\rangle$ represent zero-photon Fock state of cavity and exciton state of QD, one-photon Fock state of cavity and ground state of QD, one-photon Fock state of cavity and exciton state of QD, and two-photon Fock state of cavity and ground state of QD respectively. The couplings of employed CW laser field to the relevant states in joint basis ($|0,g\rangle \leftrightarrow |0,e\rangle$ and $|1,g\rangle \leftrightarrow |1,e\rangle$) are represented by the solid blue arrows. Similarly, the couplings of the cavity mode to the relevant states in joint basis ($|0,e\rangle \leftrightarrow |1,g\rangle$ and $|1,e\rangle \leftrightarrow |2,g\rangle$) are represented by solid red arrows. The decay of cavity Fock states ($|2,g\rangle \to |1,g\rangle, |1,e\rangle \to |0,e\rangle,$ and $|1,g\rangle \to |0,g\rangle$) with decay rate $\kappa$ are represented by dashed green arrows and the decay of exciton state ($|1,e\rangle \to |1,g\rangle$ and $|0,e\rangle \to |0,g\rangle$) with natural decay rate $\gamma$ are represented by dashed dotted black arrows.

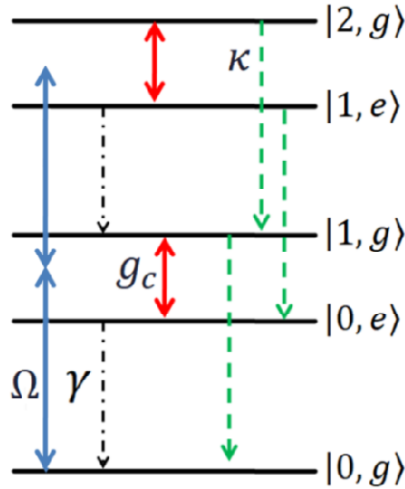

Fig. 1 (color online) Schematic of the quantum states of the QD-cavity system. The relevant couplings of CW laser field and cavity mode are represented by solid arrows and decay of QD-cavity system is represented by dashed arrows. The physical meaning of the used notations is described in the text.

### B. Effective polaron master equation theory

In contrast to the real atoms, the exciton states of QDs are unavoidably coupled to their phonon modes. For a typical coherently driven InGaAs/GaAs QDs, the coupling of longitudinal acoustic (LA)-phonon bath is found to be quite dominant and influential compared to the optical-phonon bath, leading to the excitation induced incoherent scattering and dephasing of exciton and biexciton states of QDs [40, 41]. Several theoretical methods such as path integral, variational method and polaron transformation are utilized for accurately incorporating the effect of exciton-phonon coupling depending upon the various concerned parameters like Rabi frequency, cavity coupling strength, and temperature of the phonon-bath [42-44]. In this paper, we chose to employ a polaron transformation based equation, the so called effective polaron master equation (EPME). Under suitable parameter regime, this equation is quite known for providing the accurate results, much faster computation, and deeper understanding of the phonon mediated processes through the analytical form of phonon-induced incoherent scattering rates. It is worthwhile to mention here that the EPME gives the accurate results in the regime of $\Omega^{-1}$ and $g^{-1}$ being much larger than the phonon correlation time ($\tau_{ph} \approx 2\ ps$) or when the detunings ($\Delta_{xl}, \Delta_{cx}$) are much larger than $\Omega$ and $g_c$ [45]. The EPME for our QD-cavity system shown in Fig. 1 is given below. The procedural details for deriving the EPME can be found in Refs. [44 - 46].

$$\frac{d\rho(t)}{dt} =$$

$$-\frac{i}{\hbar}[H_s, \rho(t)] + \frac{\Gamma_{ph}^{\sigma^+}}{2}\mathscr{L}[\sigma^+]\rho(t) + \frac{\Gamma_{ph}^{\sigma^-}}{2}\mathscr{L}[\sigma^-]\rho(t) + \frac{\Gamma_{ph}^{a^\dagger\sigma^-}}{2}\mathscr{L}[a^\dagger\sigma^-]\rho(t) + \frac{\Gamma_{ph}^{\sigma^+ a}}{2}\mathscr{L}[\sigma^+ a]\rho(t) + L[\rho(t)],\qquad(1)$$

The Hamiltonian of the QD-cavity system reads as

$$H_s = \hbar\Delta_{xl}\sigma^+\sigma^- + \hbar\Delta_{cl}a^\dagger a + \langle B\rangle\left(\frac{\hbar\Omega}{2}(\sigma^+ + \sigma^-) + \hbar g_c(\sigma^+ a + a^\dagger \sigma^-)\right),\qquad(2)$$

where $\Delta_{xl} = \omega_x - \omega_l$ is the detuning of CW laser field with respect to exciton state, while $\Delta_{cl} = \omega_c - \omega_l$ is the detuning of laser field with respect to cavity mode. The operators $\sigma^+ = |e\rangle\langle g|$ and $\sigma^- = |e\rangle\langle g|$ are the raising and lowering operators, $a^\dagger$ and $a$ are the one-photon creation and annihilation operators. The thermally averaged phonon displacement operators, $\langle B\rangle$, is defined as $\langle B\rangle = exp\left[-\frac{1}{2}\int_0^\infty d\omega \frac{J(\omega)}{\omega^2} coth\left(\frac{\hbar\omega}{2K_B T}\right)\right]$, where $T$ represents the temperature of phonon-bath. The phonon spectral function, $j(\omega)$, quantifies the exciton-phonon coupling and defined as $j(\omega) = \alpha_P \omega^3 exp\left(-\frac{\omega^2}{2\omega_b^2}\right)$. Here $\alpha_P$ represents the strength of exciton-phonon coupling and $\omega_b$ represents the cutoff frequency of phonon-bath. The Rabi frequency in terms of exciton electric dipole moment, $\mu$, electric field of CW laser, $E$, is defined as $\Omega = \mu E/\hbar$, $g_c$ is the QD-cavity coupling strength. The operators $\Gamma_{ph}^{\sigma^+}$ and $\Gamma_{ph}^{\sigma^-}$ represent the phonon-induced incoherent excitation and de-excitation rates of exciton state, while $\Gamma_{ph}^{a^\dagger\sigma^-}$ and $\Gamma_{ph}^{\sigma^+ a}$ represent the rates of phonon-induced creation of cavity photon accompanied by the decay of the exciton state and annihilation of the cavity photon accompanied by the excitation of the exciton state, respectively. These phonon-induced incoherent rates are given as

$$\Gamma_{ph}^{\sigma^+/\sigma^-} = \frac{\Omega_R^2}{2} Re\left[\int_0^\infty d\tau e^{\pm i\Delta_{lx}\tau}\left(e^{\phi(\tau)} - 1\right)\right] \qquad 3(a)$$

$$\Gamma_{ph}^{\sigma^+ a/a^\dagger \sigma^-} = 2g_R^2 Re\left[\int_0^\infty d\tau e^{\pm i\Delta_{cx}\tau}\left(e^{\phi(\tau)} - 1\right)\right], \qquad 3(b)$$

where, $\Omega_R = \langle B\rangle\Omega$ and $g_R = \langle B\rangle g_c$ are the phonon-renormalized Rabi frequency and phonon-renormalized cavity coupling strength, respectively. The value of mean phonon displacement, $\langle B\rangle$, is 0.91 for temperature $T = 4\ K$. The term, $\phi(\tau)$ represents the phonon phase, which is given as $\phi(\tau) = \int_0^\infty d\omega \frac{j(\omega)}{\omega^2}\left[coth\left(\frac{\hbar\omega}{2K_B T}\right) cos(\omega\tau) - i sin(\omega\tau)\right]$ in the continuum limit. The last term, $L[\rho(t)]$, in Eq. 1 is added phenomenologically for incorporating the natural radiative decay, $\gamma$ pure dephasing, $\gamma'$ rates of exciton state along with the cavity decay rate, $\kappa$. It reads as

$$L[\rho(t)] = \frac{\gamma}{2}\mathscr{L}[\sigma^-]\rho(t) + \frac{\gamma'}{2}\mathscr{L}[\sigma^+\sigma^-]\rho(t) + \frac{\kappa}{2}\mathscr{L}[a]\rho(t).$$

### III. Numerical Results and Discussions

We investigate the squeezing of cavity field by analyzing the variance of the quadratures, $X_\theta = \frac{1}{2}(a^\dagger e^{i\theta} + a e^{-i\theta})$, where $\theta$ is an adjustable phase. Furthermore, we define the variance and fluctuation of $a$ with respect to the steady state value $\langle a \rangle$ as: $\langle \Delta X_\theta^2 \rangle = \langle (X_\theta - \langle X_\theta \rangle)^2 \rangle$ and $\Delta a = a - \langle a \rangle$, respectively. The normally ordered quadrature variance for $\theta = 0$ reads as $\langle :\Delta X^2: \rangle = \frac{1}{2}[\langle \Delta a \Delta a^\dagger \rangle + Real(\Delta a^2)]$. For a vacuum or coherent state, $\langle \Delta a \Delta a^\dagger \rangle = \Delta a^2 = 0$, thus the normally ordered variance is equal to the shot noise, that is $\langle :\Delta X^2: \rangle = 0$. Therefore, the state of the cavity field will be quadrature squeezed if the normally ordered variance becomes negative, $\langle :\Delta X^2: \rangle < 0$, corresponding to a noise reduction below vacuum level. The normally ordered variance in terms of the expectation values of cavity operators reads as:

$$\langle :\Delta X^2: \rangle = \frac{1}{2}[\langle a^\dagger a \rangle - \langle a \rangle \langle a^\dagger \rangle + Re(\langle a^2 \rangle - \langle a \rangle^2)] \qquad (4)$$

The squeezing of cavity field is investigated by numerically solving Eq. 1 for calculating the expectation values of involved operators in Eq. 4, via $\langle O \rangle = Tr(O\rho)$, where $O$ is an arbitrary operator. In simulation, we use the typical values of QD parameters as: $\gamma = 2\,\mu eV$, $\alpha_P/(2\pi)^2 = 0.06\,ps^2$, $\gamma' = 0.5\,\mu eV$ and $\omega_b = 1\,meV$ [40, 41]. The values of Rabi frequency, $\Omega$ and cavity coupling strength, $g_c$ are appropriately chosen to be $55\,\mu eV$ and $82.4\,\mu eV$, respectively. It is to be noted that for these chosen values, $\Omega^{-1}$ and $g_c^{-1}$ turn out to be $75\,ps$ and $50\,ps$, respectively, which are much greater than the phonon correlation time ($\tau_{ph} \approx 2\,ps$). Therefore in this work, the validity condition of EPME is clearly satisfied.

## A. Evolution of phonon-induced incoherent rates as a function of laser-exciton and cavity-exciton detunings

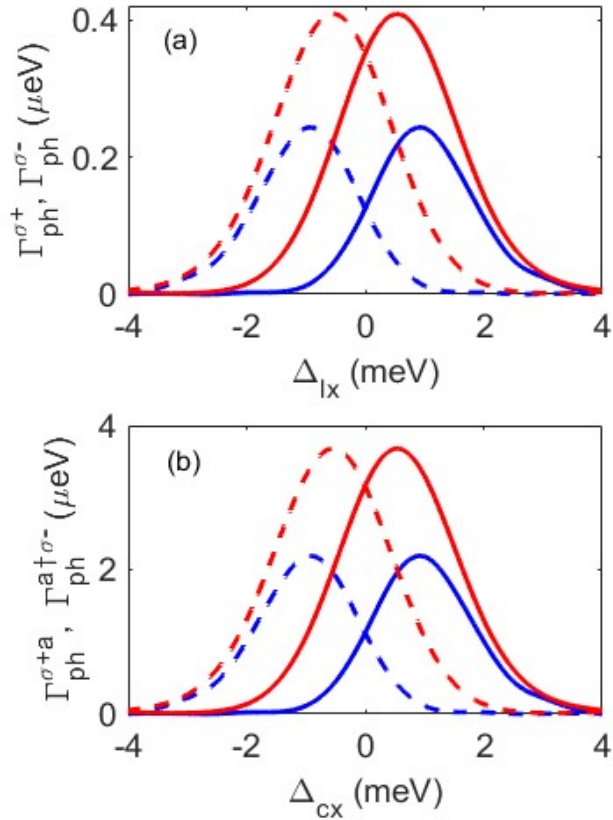

Fig. 2 (color online) Evolution of phonon-induced incoherent rates as function of detuning for a fixed value of Rabi frequency, $\Omega_R = 50\,\mu eV$, and cavity coupling strength, $g_R = 75\,\mu eV$ for the two cases: (a) $\Gamma_{ph}^{\sigma^-}$ (dashed blue line), $\Gamma_{ph}^{\sigma^+}$ (solid blue line) at $T = 4\,K$, and $\Gamma_{ph}^{\sigma^-}$ (dashed red line), $\Gamma_{ph}^{\sigma^+}$ (solid red line) at $T = 10\,K$ and (b) $\Gamma_{ph}^{a^\dagger \sigma^-}$ (dashed blue line), $\Gamma_{ph}^{\sigma^+ a}$ (solid blue line), at $T = 4\,K$ and $\Gamma_{ph}^{a^\dagger \sigma^-}$ (dashed red line), $\Gamma_{ph}^{\sigma^+ a}$ (solid red line) at $T = 10\,K$.

In Fig. 2 (a), we show the evolution of phonon-induced incoherent rates, $\Gamma_{ph}^{\sigma^+}$ and $\Gamma_{ph}^{\sigma^-}$ as a function of laser-exciton detuning for a fixed value of Rabi frequency, $\Omega_R = 50\ \mu eV$, at two different temperatures of phonon-bath, $T = 4\ and\ 10\ K$. It can be observed that the magnitude of both $\Gamma_{ph}^{\sigma^+}$ and $\Gamma_{ph}^{\sigma^-}$ is larger at $10\ K$ in comparison with $\Gamma_{ph}^{\sigma^+}$ and $\Gamma_{ph}^{\sigma^-}$ at $4\ K$. This is simply due to the fact that at increased phonon-bath temperatures, the phonon emission and absorption processes become more pronounced. Similarly in Fig. 2(b), we show the evolution of phonon-induced incoherent rates, $\Gamma_{ph}^{a^\dagger \sigma^-}$ and $\Gamma_{ph}^{\sigma^+ a}$ as a function of cavity-exciton detuning for a fixed value of cavity coupling strength, $g_R = 75\ \mu eV$, at two different temperatures of phonon-bath, $T = 4\ and\ 10\ K$. At phonon-bath temperature, $T = 10\ K$, $\Gamma_{ph}^{a^\dagger \sigma^-}$ and $\Gamma_{ph}^{\sigma^+ a}$ follow the same trend of increased magnitude as followed by $\Gamma_{ph}^{\sigma^+}$ and $\Gamma_{ph}^{\sigma^-}$ in Fig. 2(a). We also find, although not shown here, that the values of phonon-induced incoherent rates increases with respect to the increase in the phonon-bath temperature.

## B. Evolution of the exciton coherence, variance and cavity operators as a function of cavity detuning

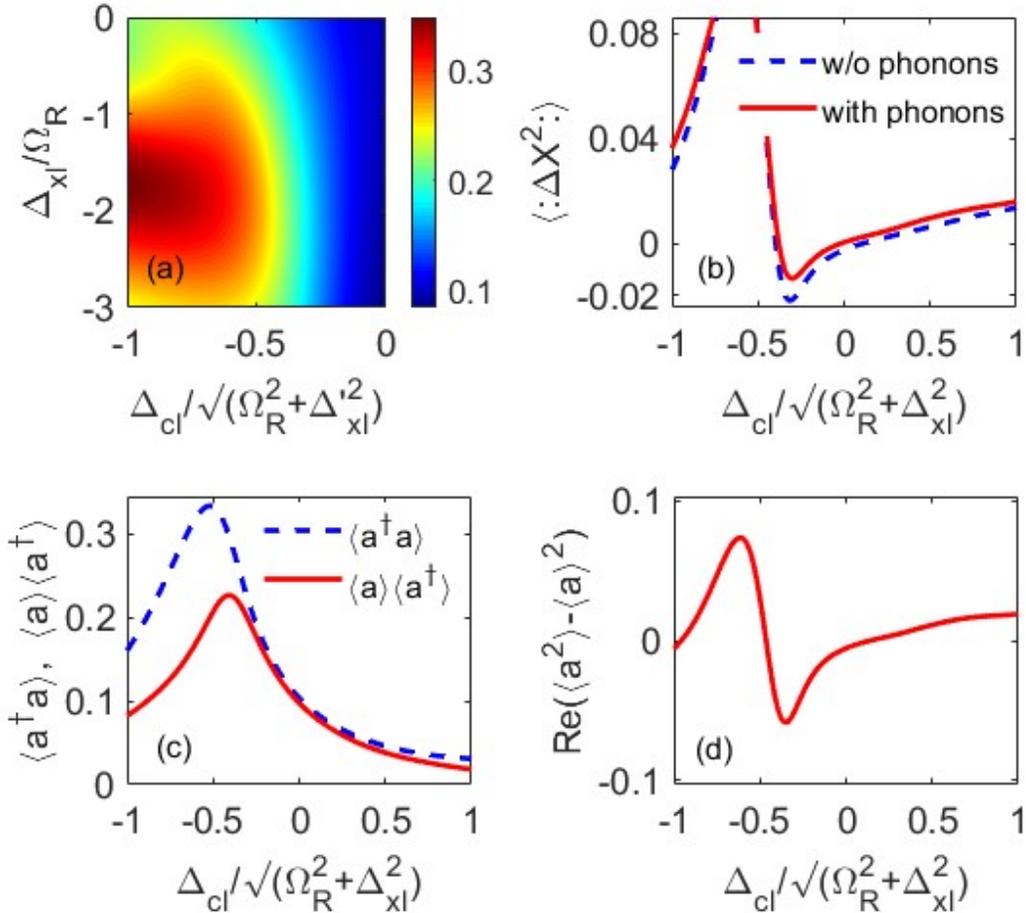

Fig. 3 (a) (color online) Contour plot of the exciton coherence, $|\langle \sigma^- \rangle|$ as a function of $\Delta_{cl}$ and $\Delta_{xl}$, (b) Evolution of the variance as a function of $\Delta_{cl}$, without including the exciton-phonon coupling (dased blue line) and with exciton-phonon coupling (solid red line), (c) Evolution of $\langle a^\dagger a \rangle$ (dashed blue line) and $\langle a \rangle \langle a^\dagger \rangle$ (solid red line) as a function of $\Delta_{cl}$, and (d) Evolution of $Re(\langle a^2 \rangle - \langle a \rangle^2)$ as a function of $\Delta_{cl}$. The values of the concerned parameters are taken as: phonon renormalized Rabi frequency, $\Omega_R = 50\ \mu eV$, cavity coupling strength, $g_R = 1.5\Omega_R$, cavity decay rate, $\kappa = 0.9\Omega_R$, and phonon-bath temperature, $T = 4\ K$.

We now investigate the appropriate parameters of the driven QD-cavity system such as Rabi frequency, detunings, cavity coupling strength, and cavity decay rate for obtaining the maximum coherence between ground and exciton state of QD. Here on wards, we termed it as the exciton coherence. The importance of the exciton coherence will be clear through the subsequent text and figures. In Fig. 3(a), we show the contour plot of exciton coherence as a function of $\Delta_{cl}$ and $\Delta_{xl}$ at the fixed values of $\Omega_R = 50\ \mu eV$ and $\Delta'_{xl} = 50\ \mu eV$. It can be observed that the coherence is maximum at $\Delta_{xl} = -1.5\Omega_R$ for $\Delta_{cl} = -\sqrt{\Omega_R^2 + \Delta'^2_{xl}}$. Next in Fig. 3(b), we show the evolution of variance, without exciton-phonon coupling (dashed blue line) and with exciton-phonon coupling (solid red line) as a function of $\Delta_{cl}$ for $\Delta_{xl} = -1.5\Omega_R$. It can be observed that value of variance becomes negative at $\Delta_{cl} = -0.3\sqrt{\Omega_R^2 + \Delta_{xl}^2}$ in both the cases, without and with exciton-phonon coupling. The negativity of variance clearly reflects the squeezing of the cavity field. Note that the magnitude of variance is slightly less in case of exciton-phonon coupling due to the phonon-induced incoherent rates, $\Gamma_{ph}^{\sigma^+}$, $\Gamma_{ph}^{\sigma^-}$, $\Gamma_{ph}^{a^\dagger \sigma^-}$, and $\Gamma_{ph}^{\sigma^+ a}$.

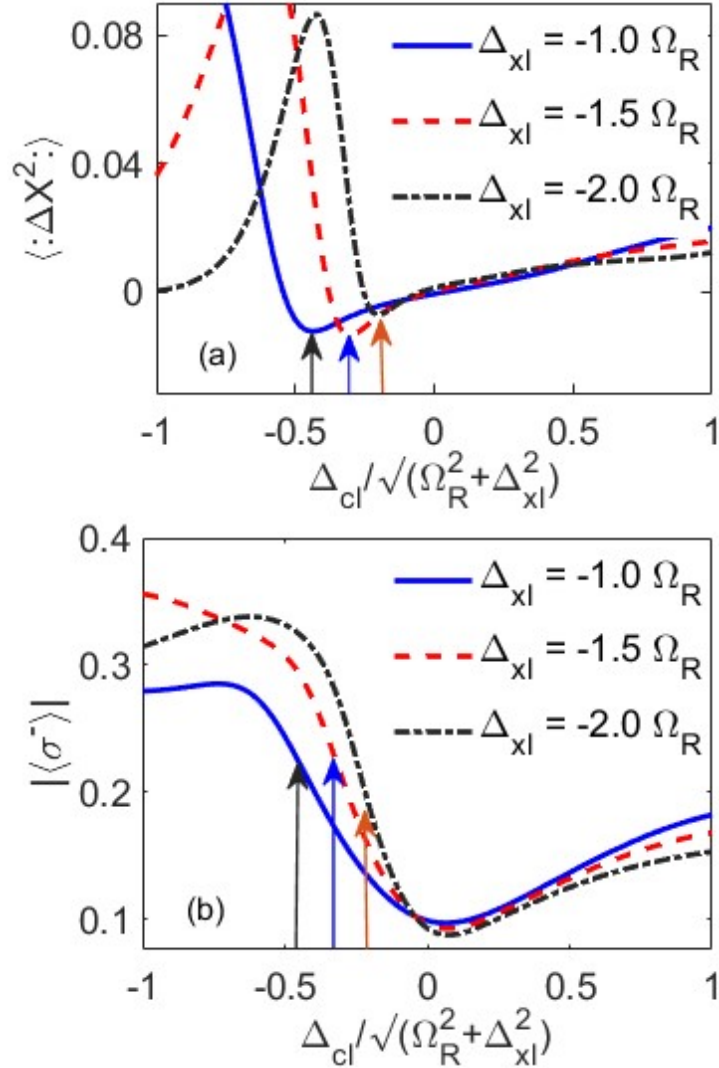

Fig. 4 (a) (color online) Evolution of the variance as a function of $\Delta_{cl}$ at three different values of $\Delta_{xl}$, and (b) Evolution of the exciton coherence as a function of $\Delta_{cl}$ at the same three values of $\Delta_{xl}$ as shown in (a). The value of other necessary parmeters are taken as: phonon renormalized Rabi frequency, $\Omega_R = 50\ \mu eV$, cavity coupling strength, $g_R = 1.5\Omega_R$, cavity decay rate, $\kappa = 0.9\Omega_R$, and phonon-bath temperature, $T = 4\ K$.

Next in Fig. 3(c) and 3(d), we show the evolution of various terms appeared in Eq. (4) for explicitly demonstrating their cavity-laser detuning dependent profiles and hence to clarify their roles in the negativity of the variance. In Fig. 3(c), we show the evolution of $\langle a^\dagger a \rangle$ and $\langle a \rangle \langle a^\dagger \rangle$ as a function of $\Delta_{cl}$. It is clear that the difference $[\langle a^\dagger a \rangle - \langle a \rangle \langle a^\dagger \rangle]$ never takes a negative value for any value of $\Delta_{cl}$, however it is minimum at $\Delta_{cl} = -0.3\sqrt{\Omega_R^2 + \Delta_{xl}^2}$. It can be observed from 3(d) that the third term of Eq. 4, $Re(\langle a^2 \rangle - \langle a \rangle^2)$ becomes maximally negative at $\Delta_{cl} = -0.3\sqrt{\Omega_R^2 + \Delta_{xl}^2}$, resulting the negative variance [see Fig. 3(b)] and therefore reflects the squeezing in the cavity field. Next, we show the evolution of the variance and compare it with exciton coherence as a function of $\Delta_{cl}$ at the three different values of $\Delta_{xl}$ in Fig. 4 for illustrating and clarifying the role of exciton coherence. It can be obseved from Fig. 4(a) that the negative value of variance is greater for $\Delta_{xl} = -1.5\Omega_R$ at $\Delta_{cl} = -0.3\sqrt{\Omega_R^2 + \Delta_{xl}^2}$ as compared to $\Delta_{xl} = -1.0\Omega_R$ at $\Delta_{cl} = -0.44\sqrt{\Omega_R^2 + \Delta_{xl}^2}$ and $\Delta_{xl} = -2.0\Omega_R$ at $\Delta_{cl} = -0.2\sqrt{\Omega_R^2 + \Delta_{xl}^2}$ as highlighted by blue arrow. We note that this is due to the larger value of coherence for $\Delta_{xl} = -1.5\Omega_R$ at $\Delta_{cl} = -0.3\sqrt{\Omega_R^2 + \Delta_{xl}^2}$ as can be seen in Fig. 4(b) which is again highlighted by blue arrow. We further derive an equaion of motion of $\langle a \rangle$ as: $\frac{d\langle a \rangle}{dt} = -i(\Delta_{cl} - i\kappa/2) - ig_R \langle \sigma^- \rangle$, for explicitely showing a connection between the variance and exciton coherence. Under steady state limit, $\langle a \rangle$ is given as: $\langle a \rangle = -\frac{g_R}{(\Delta_{cl} - i\kappa/2)} \langle \sigma^- \rangle$. From this relation, it is now explicitly clear that for the optimized values of cavity coupling strength, $g_R = 1.5\Omega_R$, cavity decay rate, $\kappa = 0.9\Omega_R$, the value of $\langle a \rangle$ varies as a fucntion of $\Delta_{cl}$ with respect to $\langle \sigma^- \rangle$. That's why, $\langle a \rangle \langle a^\dagger \rangle$ in Fig. 3(c) and $Re(\langle a^2 \rangle - \langle a \rangle^2)$ in Fig. 3(d) show the strong dependence on $\Delta_{cl}$. It should be mentioned that we have not included the phonon induced incoherent processes for deducing a relation between $\langle a \rangle$ and $\langle \sigma^- \rangle$ for the sake of simplicity. Now, with the aid of Fig. 4 and the relation between $\langle a \rangle$ and $\langle \sigma^- \rangle$ along with with Eq. 4, it is clear that the variance of cavity quadrature depends upon the exciton coherence $\langle \sigma^- \rangle$ through $\langle a \rangle$.

## C. Occupation probabilities and coherences of the Fock states of cavity mode

For illustrating that the exciton coherence faciliates the formation of coherences between the Fock states of the cavity mode, we now investigate the steady state occupation probabilities of various Fock states of the cavity mode and coherences between them. Firstly, in Fig. 5, we show the occupation probabilities of the various Fock states at $\Delta_{cl} = -0.3\sqrt{\Omega_R^2 + \Delta_{xl}^2}$ for clarifying the role of multi-photon transistions in the sqeezing of cavity field.

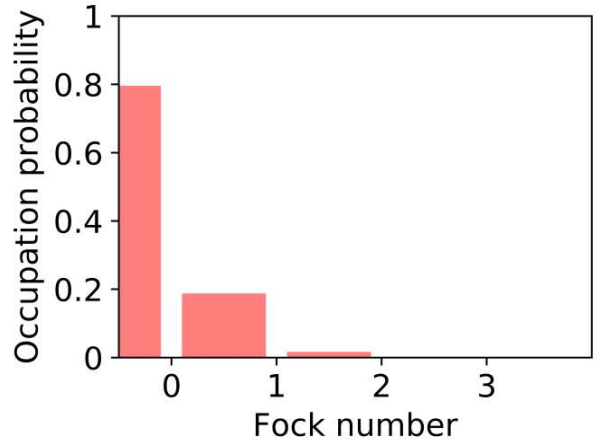

Fig. 5 (color online) Occupation probability of different Fock states of the cavity mode at $\Delta_{cl} = -0.3\sqrt{\Omega_R^2 + \Delta_{xl}^2}$ for $\Omega_R = 50\ \mu eV$, cavity coupling strength, $g_R = 1.5\Omega_R$, cavity decay rate, $\kappa = 0.9\Omega_R$, and phonon-bath temperature, $T = 4\ K$.

The occupation probabilities of the cavity mode Fock states are numerically calculated by tracing out the QD degrees of freedom using quantum optics toolbox in Python (QuTiP) [47]. It can be observed from Fig. 5 that the occupation probabilities of zero-, one-, and two-photon Fock states are 0.79, 0.20, and 0.01, respectively. The non-zero but negligibly small occupation probability of two-photon Fock state arises due to the two-photon transition, $(|0,g\rangle \to |2,g\rangle)$ via $|0,e\rangle$, $|1,g\rangle$, and $|1,e\rangle$, along with the decay via cavity ($\kappa$) and free space ($\gamma$) modes as can be understood from Fig. 1. Furthermore, it can clearly be noted that the occupation probability of one-photon Fock state is much greater than the two-photon Fock state. This is due to the fact that one-photon Fock state is populated by both, one-photon transition $(|0,g\rangle \to |1,g\rangle)$ via $|0,e\rangle$ and two-photon transition $(|0,g\rangle \to |1,e\rangle)$ via $|0,e\rangle$ and $|1,g\rangle$, together with cavity decay $(|2,g\rangle \to |1,g\rangle)$ as shown in Fig.1.

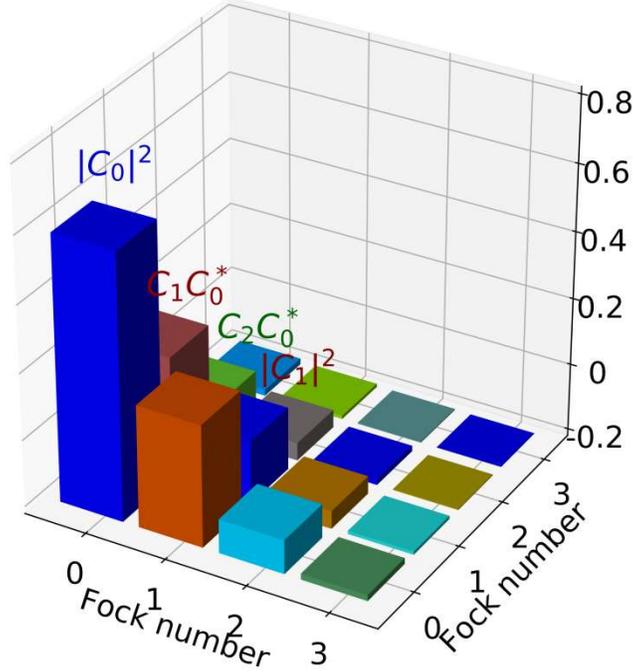

Fig. 6 (color online) Density matrix elements of the cavity mode at $\Delta_{cl} = -0.3\sqrt{\Omega_R^2 + \Delta_{xl}^2}$ for $\Omega_R = 50\ \mu eV$, cavity coupling strength, $g_R = 1.5\Omega_R$, cavity decay rate, $\kappa = 0.9\Omega_R$, and phonon-bath temperature, $T = 4\ K$.

After having clearly demonstrated that how the variance depends upon exciton coherence, $\langle \sigma^- \rangle$ through an explicit relation between $\langle a \rangle$ and $\langle \sigma^- \rangle$ and also how the dynamics of all populated Fock states takes place through Fig. 5, we now make an attempt to show that the exciton coherence results in the creation of subtantial coherence between zero- and one-photon Fock states of the cavity mode. For explicitly clarifying and illustrating this dependence, we assume that the steady state of the cavity mode can be written as: $|\psi\rangle = \sum_{n=0}^{n=2} C_n |n\rangle$. Here, we consider only up to two-photon Fock states as higher Fock states are not occupied at all as can be observed from Fig. 4. The density matrix operator of cavity mode is given as: $\rho_{cav} = |\psi\rangle\langle\psi| = [[|C_0|^2|0\rangle\langle 0| + |C_1|^2|1\rangle\langle 1| + |C_2|^2|2\rangle\langle 2| + \{C_1 C_0^*|1\rangle\langle 0| + C_2 C_1^*|2\rangle\langle 1| + C_2 C_0^*|2\rangle\langle 0| + H.C\}]$, where density matrix elements, $|C_0|^2$, $|C_1|^2$, and $|C_2|^2$ represent the occupation probabilty of zero-, one-, and two-photon Fock states, respectively, while $C_1 C_0^*$, $C_2 C_1^*$, and $C_2 C_0^*$ represent the cohrences between zero- and one-photon, one- and two-photon, zero- and two-photon Fock states, respectively.

We show the histogram of various density matrix elements of the cavity mode in Fig. 6. The diagonal blue bars represent the occupation probabilities of the Fock states, while off-diagonal bars represent the coherences between the relevant Fock states. In Fig. 6, we have highlighted especially the bars representing the elements having the significant values of occupation probabilities viz. $|C_0|^2$ and $|C_1|^2$ and significant values of coherences viz. $C_1 C_0^*$ and $C_2 C_0^*$. The signifiance of coherences between Fock states becomes very clear and evident as soon we deduce the expressions of the expectation values of annihilation operator $a$ and $a^2$ using $\langle O \rangle = Tr(O \rho_{cav})$. The expectation value of $a$ and $a^2$ are given as $\langle a \rangle = C_1 C_0^* + C_2 C_1^*$, and $\langle a^2 \rangle = C_2 C_0^*$, respectively. From these expressions, it is quite easy to comprehend that if there are no coherences ($C_1 C_0^* = C_2 C_1^* = C_2 C_0^* = 0$) then the values of $\langle a \rangle$ and $\langle a^2 \rangle$ will vanish and variance can never be negative i.e. there will be no squeezing of the cavity field [see Eq. 4]. Alternatively, it is equivalent to say that there will be no coherece between the Fock states if exciton coherence vanishes. Furthermore, it can also be noted from Fig. 6 that the coherence between zero- and one-photon is much larger compared to the one- and two-photon, and zero- and two-photon Fock states. Thereore, it is quite evident from the above disscusions to infer that that the squeezing of the cavity field in the present QD-cavity system arises due to build up of significant exciton coherence, which results in the formation of appreciable coherene between the zero- and one-photon Fock sates of the cavity mode.

D. **Evolution of the variance and exciton coherence as a function of phonon-bath temperature and density matrix elements of the cavity mode**

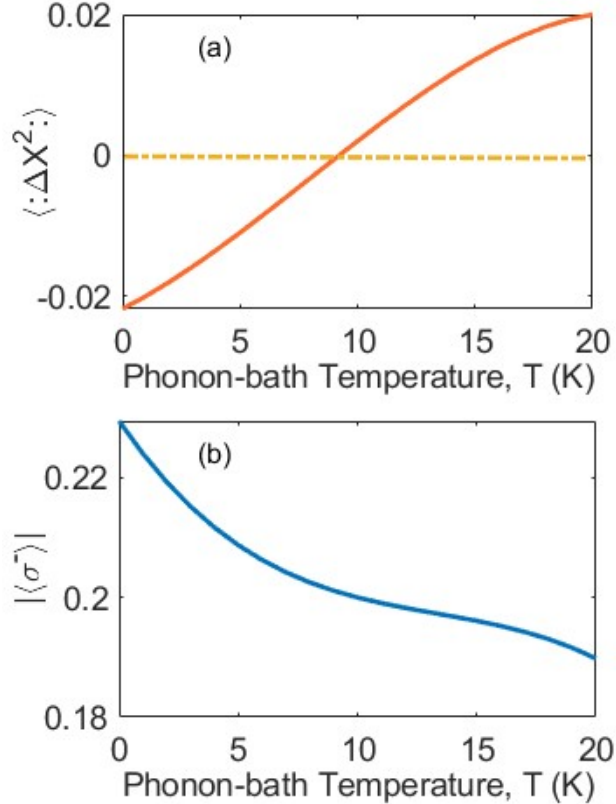

Fig. 7 (color online) Evolution of variance, $\langle : \Delta X^2 : \rangle$, and coherence, $|\langle \sigma^- \rangle|$, as a function of phonon-bath temperature at $\Delta_{cl} = -0.3\sqrt{\Omega_R^2 + \Delta_{xl}^2}$. The other parameters are same as those used in Fig. 3(b). The dashed-dotted yellow line in Fig 7 (a) reprensents a boundary above which there is no squeezing of the cavity field.

We so far have investigated the squeezing by analyzing the variance and other concerned parameters for a fixed but typical value of phonon-bath temperature, $T = 4\,K$, considered in the usual experiments, except a comparision of the variance without exciton-phonon coupling i.e.

$T = 0\ K$ in Fig. 3(b). Therefore, in Fig. 7 (a), we now show the evolution of variance for a quite wide range of fluctuation of phonon-bath temperature, ranging from 0 to 20 $K$. It can be seen that no squeezing persits beyond the phonon-bath temperature, T = 9 K. This is simply due to the increased values of phonon-induced incoherent rates, $\Gamma_{ph}^{\sigma^+}$, $\Gamma_{ph}^{\sigma^-}$, $\Gamma_{ph}^{a^\dagger \sigma^-}$, and $\Gamma_{ph}^{\sigma^+ a}$, (see Fig. 2), resulting the decreased value of exciton coherence, which eventually results in the degradation of coherences between the Fock states of the cavity mode.

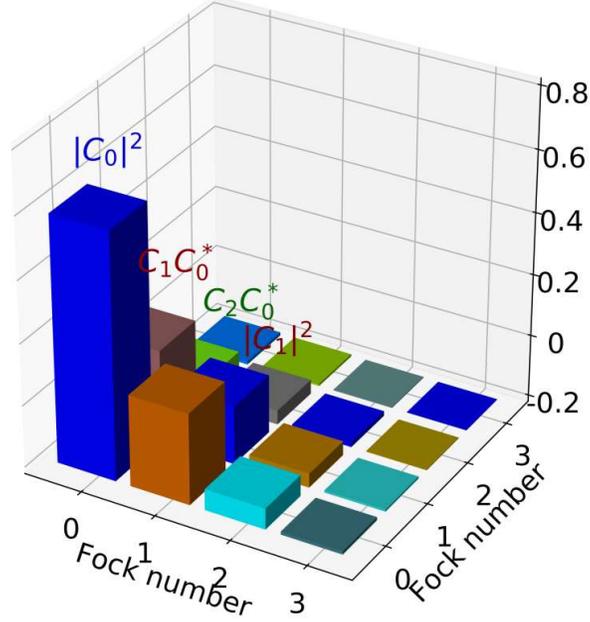

Fig. 8 (color online) Density matrix elements of the cavity mode at $\Delta_{cl} = -0.3\sqrt{\Omega_R^2 + \Delta_{xl}^2}$ for $\Omega_R = 50\ \mu eV$, cavity coupling strength, $g_R = 1.5\Omega_R$, cavity decay rate, $\kappa = 0.9\Omega_R$, and phonon-bath temperature, $T = 9\ K$.

To illustrate this, we show the evolution of exciton coherence in Fig. 7 (b) and density matrix elements of the cavity mode in Fig. 8. It can be observed from Fig. 7 (b) that the value of exciton coherence significantly decreases as the phonon-bath temperatures increases due to the increased phonon-induced incoherent rates, $\Gamma_{ph}^{\sigma^+}$, $\Gamma_{ph}^{\sigma^-}$, $\Gamma_{ph}^{a^\dagger \sigma^-}$, and $\Gamma_{ph}^{\sigma^+ a}$ as can be understood from Fig. 2. In Fig. 8, we show the density matrix elements of the cavity mode at phonon-bath temperature, $T = 9\ K$. A comparative observation of Fig. 6 and Fig. 8 reveals that the magnitude of coherences, $C_1 C_0^*$, and $C_2 C_0^*$ at 9 $K$ is smaller than compared to their values at 4 $K$. That's why no squeezing survive at phonon-bath temperature, $T = 9\ K$ and beyond.

### IV. Conclusions

We have investigated a coherently driven quantum dot coupled to a single mode pillar microcavity by employing an effective polaron master equation theory for accurately incorporating the effects of exciton-phonon coupling. We show that the squeezing of the cavity field can be realized due to the build-up of exciton coherence under appropriately chosen parameters of the QD-cavity system. We have also investigated the occupation probabilities and coherences between the Fock states of cavity mode to show that exciton coherence faciliciates the creation of appreciable coherence between zero- and one-photon Fock states of the cavity mode, which results in the squeezing of cavity field. Furthermore, the effect of exciton-phonon coupling is shown to be detrimental for the realization of squeezing, and in fact no squuzing persists for a phonon-bath temerature of greater than 9 $K$.

**Aknowledgement**

This work is partially supported by a Dr. D. S. Kothari Postdoctoral Research Fellowship, University Grant Commission, India, through Grant No.F.4-2/2006 (BSR)/PH/15-16/0077. One of the authors, Parvendra Kumar would like to thank Prof. Amarendra Kumar Sarma for the useful discussions.